\documentclass[10pt]{article}
\usepackage{amsmath,amssymb,booktabs,graphicx,geometry}
\usepackage{epigraph}
\usepackage{subcaption}
\usepackage{graphicx}
\usepackage{caption}
\title{Fast Times, Slow Times: Timescale Separation in Financial Timeseries Data}
\author{Jan Rosenzweig}
\date{}

\begin{document}

\maketitle

\begin{abstract}
Financial time series exhibit multiscale behavior, with interaction between multiple processes operating on different timescales. This paper introduces a method for separating these processes using variance and tail stationarity criteria, framed as generalized eigenvalue problems. The approach allows for the identification of slow and fast components in asset returns and prices, with applications to parameter drift, mean reversion, and tail risk management. Empirical examples using currencies, equity ETFs and treasury yields illustrate the practical utility of the method.
\end{abstract}

\section*{}


\section{Stationarity in Finance}
The study of multiscale processes is well-established in physics, with notable examples ranging from fluid dynamics to protein folding, where fast and slow dynamics interact \cite{hinch}. 

The typical physical example involves fast oscillations inside a slowly moving envelope, where the cumulative effect of fast oscillations drives changes in the envelope  \cite{hinch,protein}.

In finance, we can see multiple examples of similar multiscale behavior. For example, the S\&P 500 exhibits mean reversion over miliseconds (due to low-latency arbitrage between e-mini S\&P futures, ETFs and stock baskets) and over years (relative to gold or bonds). 

A notable example of feedback from high frequency to macro is of course the Flash Crash of 2010, where multiple feedback loops in high and mid frequency trading lead to the S\&P 500 losing 9\% of its value, only to recover later \cite{flash}.

Therefore identification of such processes operating on different timescales, and the nature of their interaction, is of clear importance both to market practitioners and to economists.

The key questions in this context are:
\begin{itemize}
\item Can we separate these processes?
\item Can we estimate their relaxation timescales?
\item Can we assess stationarity in a meaningful way?
\end{itemize}

This paper addresses these questions by introducing a framework for timescale separation in financial time series, building on methods from signal processing and variational principles.


On the trading side, financial strategies rely on calibrated parameters that drift over time, leading to strategy degradation.  

Two types of stationarity are relevant in this context:
\begin{itemize}
\item \textbf{Stationarity of Returns}: Implies stable distribution of returns, leading to stable parameters.
\item \textbf{Stationarity of  Prices}: Implies mean reversion in price levels.
\end{itemize}

The same mathematical tools can be applied to both, though their interpretations differ.

\section{Types of Stationarity}

The textbook definition of stationary processes involves the stationarity of the underlying distribution.

While this definition is attractive, it does not lend itself straightforwardly to data driven analysis. Primarily, this is due to the fact that we only have a finite amount of data available, and that any estimation of the underlying distribution is therefore fuzzy and imprecise. 

Rather than aiming for the entire distribution, it is then more practical to focus on particular aspects of the distribution. 

The initial promising target is the covariance matrix, i.e. measuring the stationarity of variances and covariances in the basket. 

However non-Gaussianity of financial time series means that we have to look further than just to the Gaussian MLE, into the stationarity of the tail behaviour of the relevant timeseries.

We therefore consider two complimentary approaches to timescale separation:
\begin{itemize}
\item \textbf{Variance Timescales}: Stationarity and drift in the second moment.
\item \textbf{Tail Timescales}: Stability in higher moments, relevant for tail risk.
\end{itemize}

\subsection{Variance Timescales}

Let \( \mathbf{X}_t \) be an $n$-dimensional column vector process (e.g., prices or returns). We seek a matrix $\mathbf{W}$ of $n$-dimensional column weight vectors $\mathbf{w}_{i} $ such that the drift of variance is minimized for a fixed unit of variance:

\begin{equation}
\frac{d}{dt} \text{Var}(\mathbf{X}_t\mathbf{w}_{i}^T ) \rightarrow \min, \ i=1..n
\label{minVar}
\end{equation}
\begin{equation}
\text{Var}(\mathbf{W} \mathbf{X}_t^T) = \mathbf{1}
\label{orthoVar}
\end{equation}
Here, (\ref{minVar}) and (\ref{orthoVar}) are both instantaneous at time $t$ and hence they do not contradict each other.

This leads to the generalized eigenvalue problem:

\begin{equation}
\mathbb{E}[(d\mathbf{X}_t)^T \mathbf{X}_t + \mathbf{X}_t^T d\mathbf{X}_t] \mathbf{w} = \beta \ \mathbb{E}[\mathbf{X}_t^T \mathbf{X}_t]  \mathbf{w}
\label{cont}
\end{equation}
\begin{equation}
\mathbb{E}[\mathbf{w}_{i} \mathbf{X}_t^T \mathbf{X}_t \mathbf{w}_{j}^{T}] = \delta_{ij}
\end{equation}
where $\beta$ is the  Lagrange multiplier.

Discretising $dX_{t}$ as
\[
d\mathbf{X}_{t} \approx \frac{1}{T} \left( d\mathbf{X}_{t+T} - d\mathbf{X}_{t} \right),
\]
(\ref{cont}) becomes 
\begin{equation}
\mathbf{C}(t, T) \mathbf{w} = \lambda \  \mathbf{C}(t, 0) \mathbf{w}
\label{tICA}
\end{equation}
where $\mathbf{C}(t, T)$ is the autocovariance of $\mathbf{X}_t$ with lag \( T \),
\[
\mathbf{C}(t, T) = \frac{1}{2} \mathbb{E}[\mathbf{X}_{t+T}^T \mathbf{X}_t + \mathbf{X}^T \mathbf{X}_{t+T}]
\]
and
$$\lambda = \frac{1 + \beta}{2T}$$

The eigenvalue problem (\ref{tICA}) is known as tICA \cite{schmid}, although our derivation is different from the usual tICA derivation; the usual derivation assumes that the discretised process follows a Hidden Markov Model 
\begin{equation}
    \mathbf{X}_{t+T} = \mathbf{A} \mathbf{X}_{t} + \epsilon_{t} \label{HMM}
\end{equation}
for some unknown linear operator $\mathbf{A}$ and noise $\epsilon_{t}$. Multiplying both sides of (\ref{HMM}) by $\mathbf{X}_{t}^{T}$ and taking the expectation, this yields
$$
\mathbb{E}[\mathbf{X}_{t}^{T}\mathbf{X}_{t+T}] = \mathbf{A} \  \mathbf{C}(t, 0) 
$$
where the autocovariance term $\mathbb{E}[\mathbf{X}_{t}^{T}\mathbf{X}_{t+T}]$ is usually symmetrized to give the eigenvalue problem (\ref{tICA}).

Note that the matrix $\mathbf{C}(t, 0)$ on the right hand side of equation (\ref{tICA}) is the covariance matrix, and as such it is real, symmetric and positive definite. This makes (\ref{tICA}) straightforward to solve numerically, using e.g. \texttt{LAPACK}'s \texttt{gvd} driver or its various interfaces such as Python \texttt{scipy.linalg.eigh}.

The eigenvectors of (\ref{tICA}) correspond to the directions of fastest (slowest) decay in non-stationarity. The eigenvalues $\lambda_{i}$ are their autocorrelations, as seen from (\ref{tICA}), and are therefore in the range $[-1,1]$. They are related to decay time scales through
\begin{equation}
    t_{i} = \frac{-2T}{\lambda_i-1}
    \label{timescalesVar}
\end{equation}

\subsection{Tail Timescales}

For tail timescales, we minimize the drift of higher order moments \cite{tale}. The moment of order $2k$ is given as
$$
M_{2k} = \frac{1}{2k}\mathbb{E} (\mathbf{X}_t \mathbf{w}^T )^{2k} 
$$

and the minimization problem becomes
\begin{equation}
\frac{d}{dt}\frac{1}{2k} \mathbb{E} (\mathbf{X}_t\mathbf{w}_{i}^T )^{2k} \rightarrow \min, \ i=1..n
\label{minMom}
\end{equation}
\begin{equation}
\mathbb{E}[\mathbf{w}_{i} \mathbf{X}_t^T \mathbf{X}_t \mathbf{w}_{j}^{T}] = \delta_{ij}
\label{orthoMom}
\end{equation}

The problem (\ref{minMom}), (\ref{orthoMom}) is nonlinear and it does not reduce to a simple eigenvalue problem. 

The constrained minimum satisfies
\begin{equation}
    (2k-1) \mathbb{E}[ (\mathbf{X}_t\mathbf{w}^T )^{2k-2}(d\mathbf{X}_t\mathbf{w}^T )\mathbf{X}_t] + \mathbb{E}[ (\mathbf{X}_t\mathbf{w}^T )^{2k-1}d\mathbf{X}_t] = \beta \mathbf{w}
    \label{lagrange}
\end{equation}
\begin{equation}
\mathbf{w}^{T} \mathbf{w} = \mathbf{1}
    \label{orthoLang}
\end{equation}
where $\beta$ is the Lagrange multiplier as before. 

Multiplying the right hand side of (\ref{lagrange}) by $\mathbf{w}^T$ and using (\ref{orthoLang}), we can solve for $\beta$ as
\begin{equation}
    \beta = 2k \mathbb{E}[ (\mathbf{X}_t\mathbf{w}^T )^{2k-1}(d\mathbf{X}_t\mathbf{w}^T )] = 2k \frac{d}{dt} M_{2k}
    \label{beta}
\end{equation}

From (\ref{beta}), we get the generalization of the equivalent of $\lambda$ from the previous section as
\begin{equation}
    \lambda =  \mathbb{E}[ (\mathbf{X}_t\mathbf{w}^T )^{2k-1}(\mathbf{X}_{t+T}\mathbf{w}^T )],\label{lambda}
\end{equation}
so the eigenvalue is now the {\it tail autocorrelation} of order $k$,  i.e. the tail correlation of order $k$ between $\mathbf{X}_{t}$ and $\mathbf{X}_{t+T}$ \cite{tale}.

Plugging (\ref{beta}) into (\ref{orthoMom}), we get 
\begin{equation}
       (2k-1) \mathbb{E}[ (\mathbf{X}_t\mathbf{w}^T )^{2k-2}(d\mathbf{X}_t\mathbf{w}^T )\mathbf{X}_t] + \mathbb{E}[ (\mathbf{X}_t\mathbf{w}^T )^{2k-1}d\mathbf{X}_t] = 2k \mathbb{E}[ (\mathbf{X}_t\mathbf{w}^T )^{2k-1}(d\mathbf{X}_t\mathbf{w}^T )] \mathbf{w}
\end{equation}

Numerically, it is easier to solve (\ref{minMom}),(\ref{orthoMom})  using the Fixed Point Iteration of \cite{fastica}, known as FastICA. The iteration procedure is
$$
\textbf{w} \leftarrow \mathbb{E}[ \mathbf{X}_t d (\mathbf{w}^t\mathbf{X}_t)^{2k-1})] - (2k-1) \mathbb{E}[(\mathbf{w}^T\mathbf{X}_t)^{2k-2}) ] \mathbf{w}
$$
\begin{equation}
    \textbf{w} \leftarrow \textbf{w} / |\textbf{w}|
    \label{deflation}
\end{equation}
when iterating a single component, or 
$$
\textbf{W} \leftarrow \mathbb{E}[ \mathbf{X}_t d (\mathbf{W}^T \mathbf{X}_t)^{2k-1})] - (2k-1) \mathbb{E}[(\mathbf{W}^T \mathbf{X}_t)^{2k-2}) ] \mathbf{W}
$$
\begin{equation}
    \textbf{W} \leftarrow \left(\mathbf{W} \mathbf{X}_t^T  \mathbf{X}_t \mathbf{W}^T \right)^{-1/2} \mathbf{W}
    \label{deflation}
\end{equation}
when iterating all the components in parallel.

\section{Empirical Examples}

We have applied the time-driven decomposition to three separate baskets, namely the G10 currencies, five multifactor equity ETFs, and US Treasuries.

For G10 currencies, we extracted the five slowest components from 2006–2010 and projected them onto 2010–2025. 

For factor ETFs (IFSU, IUSZ, IUVL, IUMO, IUQA), we extracted the five slowest components over 2021–2022 and projected them through 2022–2025.

Finally, for US Treasuries, we fitted over 1994-1998, and continued over 1998-2025.

We fitted each basket with the liner tICA ($k=1$, section 2.1) and nonlinear tICA ($k=4$, section 2.2). 

\begin{figure}[h]
\centering
    \begin{subfigure}[b]{0.5\textwidth}
        \centering
        \includegraphics[width=0.8\textwidth]{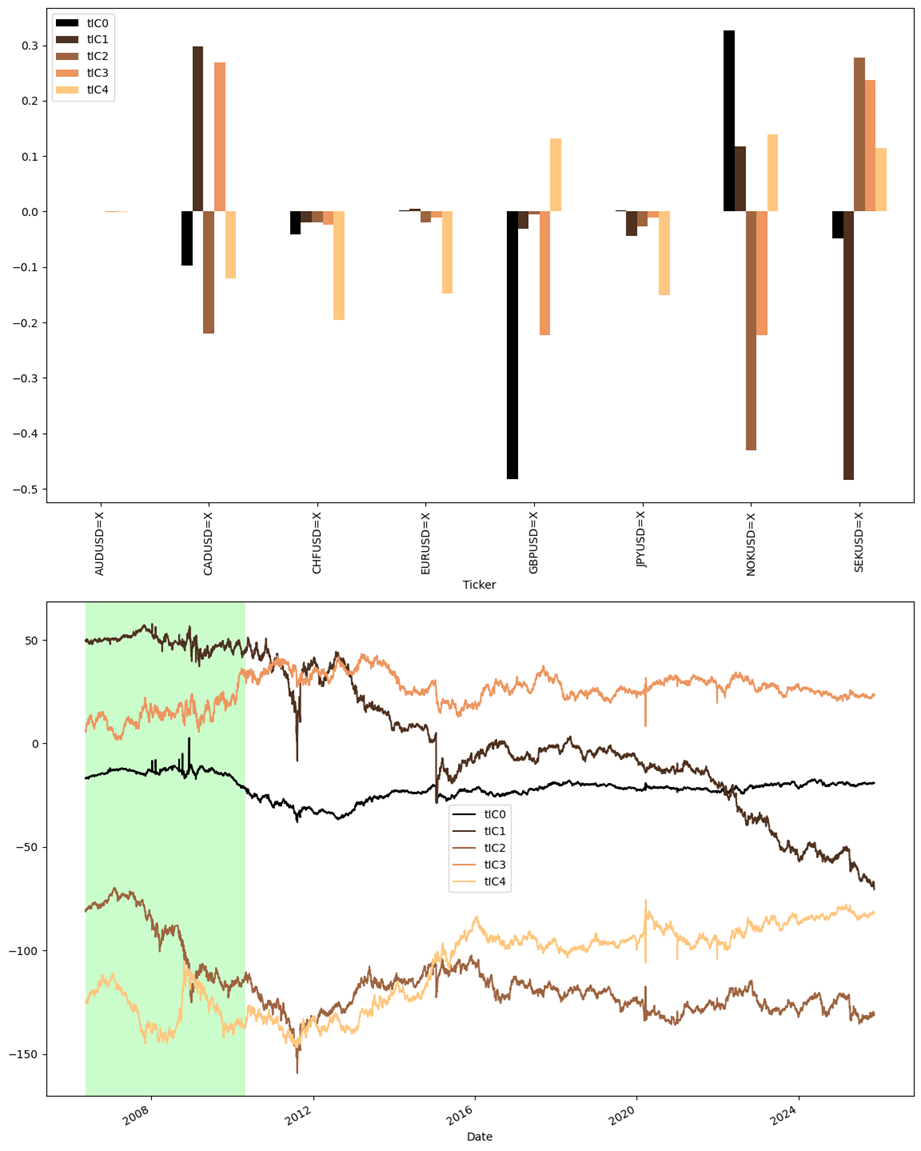}
        \caption{Linear tICA, $k=1$}
    \end{subfigure}%
    \begin{subfigure}[b]{0.5\textwidth}
        \centering
        \includegraphics[width=0.8\textwidth]{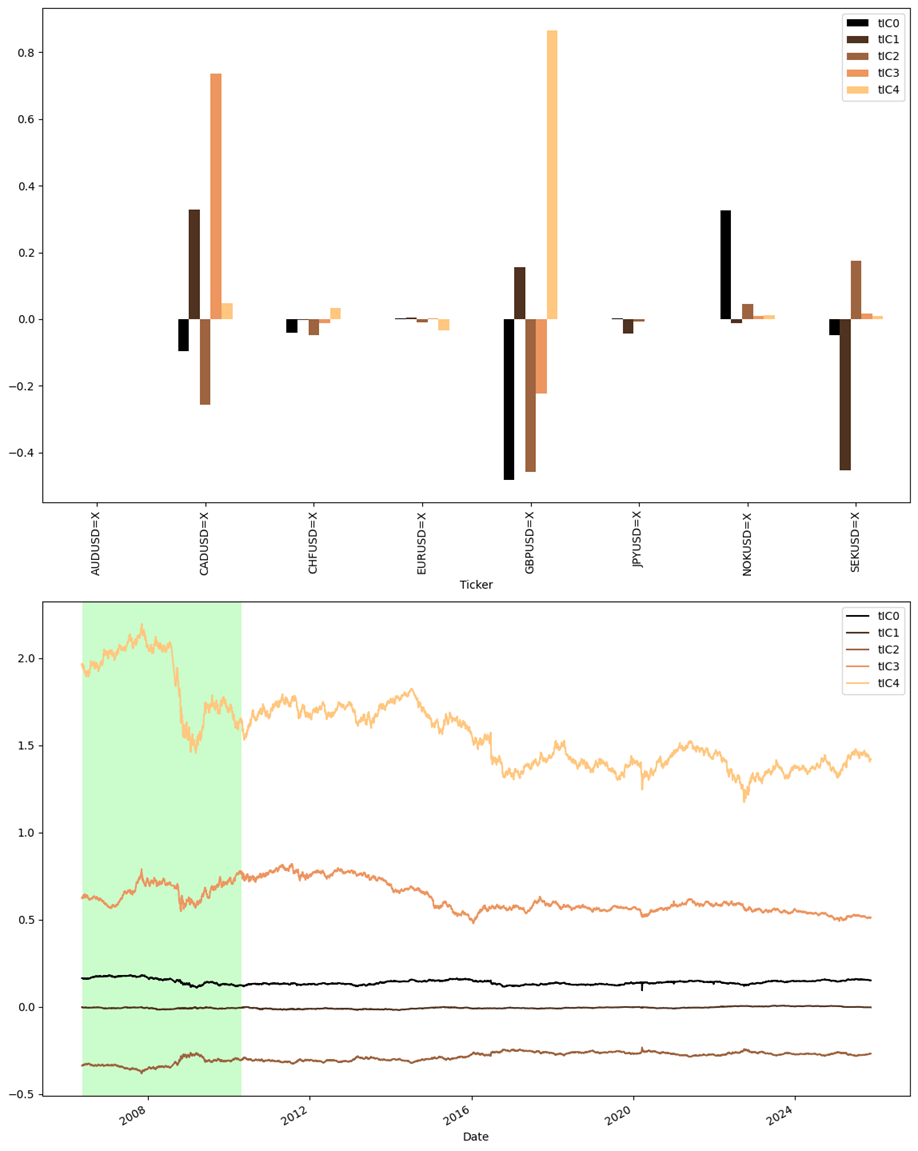}
        \caption{Nonlinear tICA, $k=4$}
    \end{subfigure}%
    \caption{G10 currencies: Blind extraction of slowest components over 2006-2010 (green background), continuation over 2010-2025 (white background). Weights are rescaled to unit gross exposure.}
\label{fig:g10}
\end{figure}

\begin{figure}[h]
\centering
    \begin{subfigure}[b]{0.5\textwidth}
        \centering
        \includegraphics[width=0.8\textwidth]{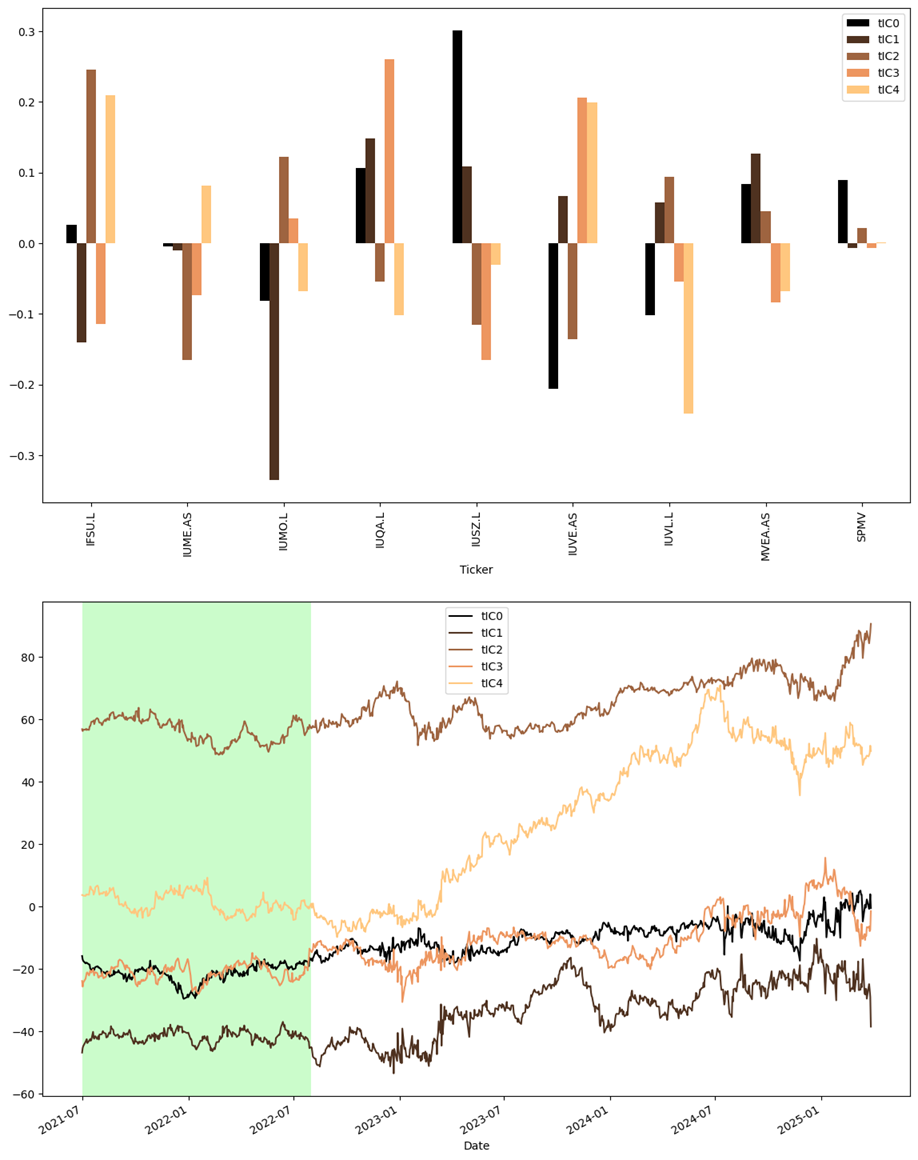}
        \caption{Linear tICA, $k=1$}
    \end{subfigure}%
    \begin{subfigure}[b]{0.5\textwidth}
        \centering
        \includegraphics[width=0.8\textwidth]{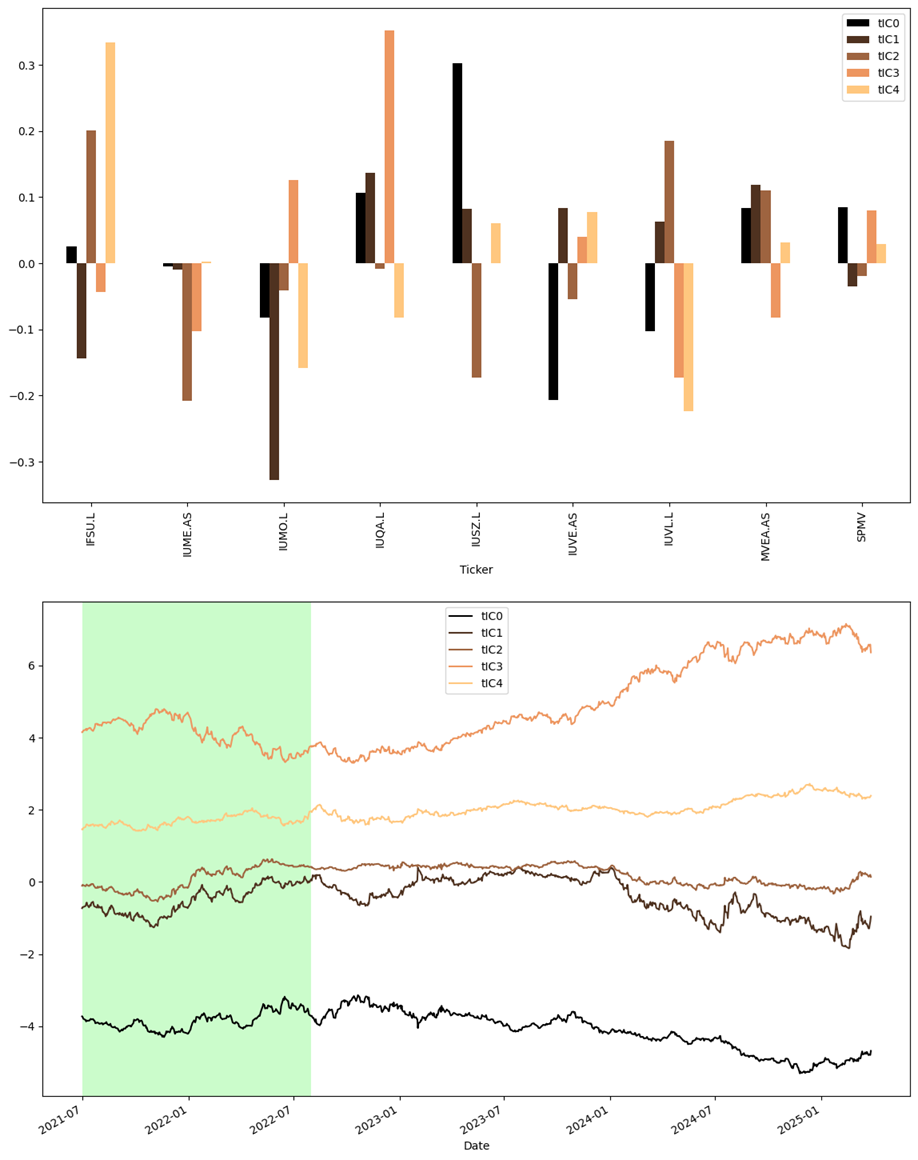}
        \caption{Nonlinear tICA, $k=4$}
    \end{subfigure}%
\caption{Factor ETF: Blind extraction of slowest components over 2021-2022 (green background), continuation over 2022-2025 (white background). Weights are rescaled to unit gross exposure.}
\label{fig:etfs}
\end{figure}

\begin{figure}[h]
\centering
    \begin{subfigure}[b]{0.5\textwidth}
        \centering
        \includegraphics[width=0.8\textwidth]{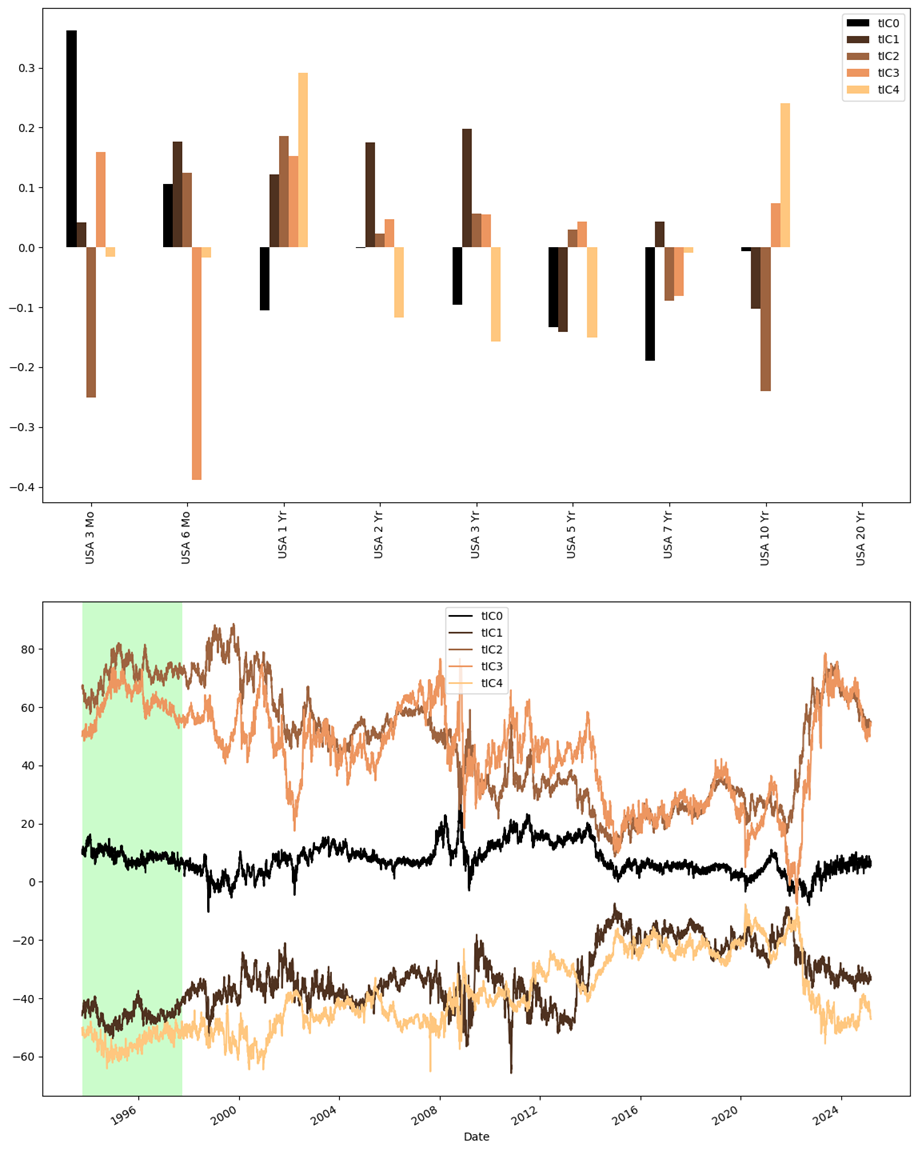}
        \caption{Linear tICA, $k=1$}
    \end{subfigure}%
    \begin{subfigure}[b]{0.5\textwidth}
        \centering
        \includegraphics[width=0.8\textwidth]{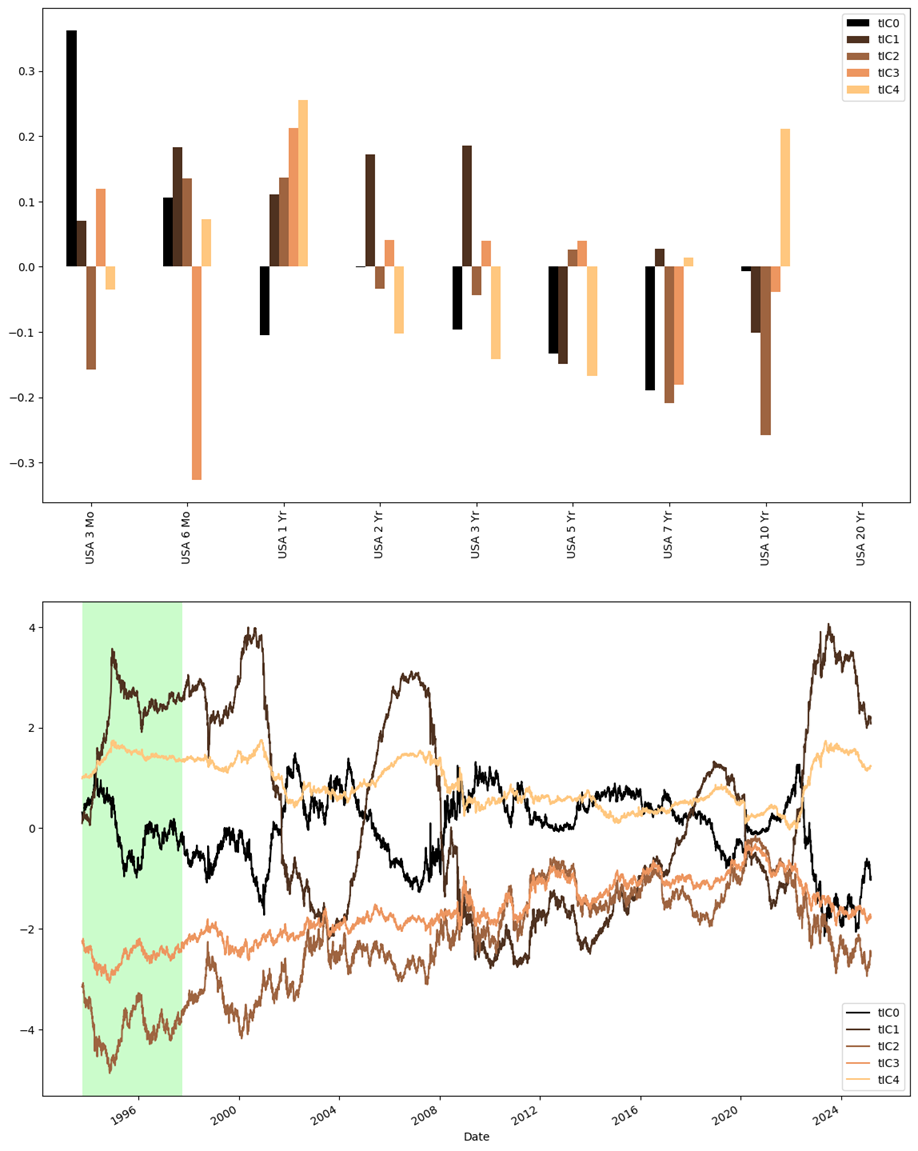}
        \caption{Nonlinear tICA, $k=4$}
    \end{subfigure}%
\caption{US Treasuries: Blind extraction of slowest components over 1994-1998 (green background), continuation over 1998-2025 (white background). Weights are rescaled to unit gross exposure.}
\label{fig:tsy}
\end{figure}

The results are shown in Figures \ref{fig:g10}, \ref{fig:etfs}, \ref{fig:tsy}.

While retaining some similarity between the weight profiles in each case, the timeseries for nonlinear tICA results are considerably different from the corresponding linear decompositions. Increasing the exponent $k$ further, and thus increasing the penalty to tails relative to volatility, did not produce further meaningful changes.

This seems to imply that there are overall two choices for timescale separation; namely linear, governed by volatility, and nonlinear, governed by the tails.

We can note that the relaxation timescales, at least for the slowest components, appear to be remarkably robust out of sample, with no visible deterioration accompanying the transition from in sample to out of sample.

Even more remarkably, where higher frequencies visibly enter the slow components such as, say, due to the global financial crisis 2008-2010 or geopolitical events in 2023, the relevant components revert back to their original, slow timescale once the relevant events pass.

This robustness does not carry over to orthogonality of components, which decays as expected out of sample.

\section{Conclusion}

Financial time series contain a hierarchy of processes: stationary, slow, and fast. Using generalized eigenvalue problems, we can separate these components, providing insight into parameter drift, mean reversion regimes and/or tail risk stability.

The method is computationally tractable and can be applied to diverse asset classes, offering a powerful tool for risk and strategy management. The conceptual framework, moving from small to large structures, mirrors approaches used in physical sciences.

The remarkable property of the methods described in this paper is their ability to persist the in-sample time scales of the selected components out of sample.

\end{document}